\documentclass[twocolumn,english,aps,pra,longbibliography,superscriptaddress]{revtex4-2}
\usepackage[T1]{fontenc}
\usepackage[latin9]{inputenc}
\usepackage{color}
\usepackage{xcolor}
\usepackage{amsmath}
\usepackage{graphicx}
\usepackage{natbib}
\usepackage{braket}
\usepackage[normalem]{ulem}
\usepackage{mathtools}

\usepackage[colorlinks=true,urlcolor=blue,citecolor=blue,linkcolor=blue]{hyperref}

\makeatother

\usepackage{babel}

\begin{document}
	
	\title{High-speed calibration and characterization of superconducting quantum processors without qubit reset}
	
	\author{M. Werninghaus}
	\email{maw@zurich.ibm.com}
	\affiliation{IBM Quantum, IBM Research -- Zurich, S\"aumerstrasse 4, 8803 R\"uschlikon, Switzerland}
	\affiliation{Department of Physics, Technical University of Munich, 85748 Garching, Germany}
	\affiliation{Theoretical  Physics,  Saarland  University,  66123  Saarbr\"ucken,  Germany}
	
	\author{D. J. Egger}
	\email{deg@zurich.ibm.com}
	\affiliation{IBM Quantum, IBM Research -- Zurich, S\"aumerstrasse 4, 8803 R\"uschlikon, Switzerland}
	\author{S. Filipp}
	\affiliation{Department of Physics, Technical University of Munich, 85748 Garching, Germany}
	\affiliation{IBM Quantum, IBM Research -- Zurich, S\"aumerstrasse 4, 8803 R\"uschlikon, Switzerland}
	
	\begin{abstract}
		To Characterize and calibrate quantum processing devices a large amount of measurement data has to be collected.
		Active qubit reset increases the speed at which data can be gathered but requires additional hardware and/or calibration.
		The experimental apparatus can, however, be operated at elevated repetition rates without reset.
		In this case, the outcome of a first measurement serves as the initial state for the next experiment. Rol. \emph{et al.} used this restless operation mode to accelerate the calibration of a single-qubit gate by measuring fixed-length sequences of Clifford gates which compose to $X$ gates [Phys. Rev. Appl. \textbf{7}, 041001 (2017)]. 
		However, we find that, when measuring pulse sequences which compose to arbitrary operations, a distortion appears in the measured data.
		Here, we extend the restless methodology by showing how to efficiently analyze restless measurements and correct distortions to achieve an identical outcome and accuracy as compared to measurements in which the superconducting qubits are reset.
		This allows us to rapidly characterize and calibrate qubits.
		We illustrate our data collection and analysis method by measuring a Rabi oscillation at a $250~\rm{kHz}$ repetition rate without any reset, for a qubit with a decay rate of $1/2\pi T_1=3~\rm{kHz}$.
		We also show that we can measure a single- and a two-qubit average gate fidelity with Randomized Benchmarking 20 and 8 times faster, respectively, than measurements that reset the qubits through $T_1$-decay.
	\end{abstract}
	
	\date{\today}
	
	\maketitle
	
	\section{Introduction}
	
	Characterizing quantum information processing devices \cite{Devoret2013} and calibrating the unitary gates is a measurement intensive task \cite{Klimov2020}.
	Quantum gates can be calibrated using error amplifying gate sequences \cite{Sheldon2016} and black-box optimization algorithms inspired from optimal control \cite{Egger2014, Kelly2014, Werninghaus2020} which are particularly measurement intensive.
	Noisy quantum processors that implement Richardson error mitigation are strongly affected by the time required to calibrate and characterize the quantum gates \cite{Temme2017, Kandala2019} as multiple copies of each gate, implemented with pulses of varying length, are needed.
	When calibrating qubit gates, for instance with error amplifying sequences \cite{Sheldon2016}, the most measurement intensive task is often determining the resulting gate fidelity which is typically done with Randomized Benchmarking (RB) \cite{Magesan2011, Magesan2012, Corcoles2013}.
	
	To guarantee a known and well defined initial state the qubits are reset to the ground state in between two consecutive measurements \cite{DiVincenzo2000}.
	The easiest way to reset a qubit is by waiting several times the decay time ($T_1$) so that the energy stored in the qubit relaxes to the environment.
	As $T_1$ times increase beyond $100~\mu\rm{s}$ \cite{Rigetti2012}, this qubit reset mechanism becomes inefficient and lengthens the time needed to acquire data.
	Qubit reset schemes, both conditional \cite{Riste2012, Govia2015} and unconditional \cite{Geerlings2013, Egger2018, Magnard2018} on the qubit state, have therefore been developed.
	However, such reset mechanisms require additional hardware and/or calibration.
	
	Qubit reset is not required to calibrate single-qubit gates since the outcome of a projective measurement \cite{Blais2004, Lupascu2007} can be used as the initial state of the next operation \cite{Rol2017}.
	In this operation mode, named \emph{restless tune-up}, data is gathered at elevated trigger rates without any qubit reset.
	This allows data to be collected at a rate of the order of $100~\rm{kHz}$, limited by the pulse lengths, when characterizing a quantum processor or calibrating the unitary gates and is essential for data intensive optimal control schemes \cite{Werninghaus2020}.
	
	Restless measurements have been used to successfully tune-up superconducting qubit gates with sequences of Clifford gates which ideally always compose to a state flip of the qubit \cite{Rol2017}.
	However, we find that, when measuring pulse sequences which compose to arbitrary operations, a distortion appears in the measured data.
	Since in restless measurements the initial state is not always the qubit ground state, the conventional data processing method that averages the measurement response does not apply \cite{Wallraff2005, Krantz2019}.
	Single-shot data can be used to resolve this issue.
	However, this requires high signal to noise ratios and the method becomes computationally expense.
	For instance, a singular-value decomposition scales quadratically with the number of IQ-points measured.
	
	In this paper we show the processing steps required to efficiently analyze restless single-shot data and how to overcome the distortions in the measurements. 
	We present our setup in Sec. \ref{sec:setup}.
	In Sec.~\ref{sec:restless_intro} we use sequences of $Id$ and $X$ gates to show the conceptual differences between restless measurements and standard measurements.
	We show, in Sec.~\ref{sec:restless_axis}, how to efficiently reconstruct a restless signal and, in Sec.~\ref{sec:bias}, how to process the data to avoid distortions caused by the different initial states; we therefore extend the usability of restless measurements to conventional tune-up and characterization.
	In Sec.~\ref{sec:rabi} we use these methods to measure a Rabi oscillation.
	Finally, in Sec.~\ref{sec:tune_up}, we show that restless measurements speed-up RB.
	We discuss our results and conclude in Sec.~\ref{sec:conclusion}.

	\section{Experimental setup\label{sec:setup}}
	\begin{figure}[!t]
		\begin{center}
			\includegraphics[width=\columnwidth]{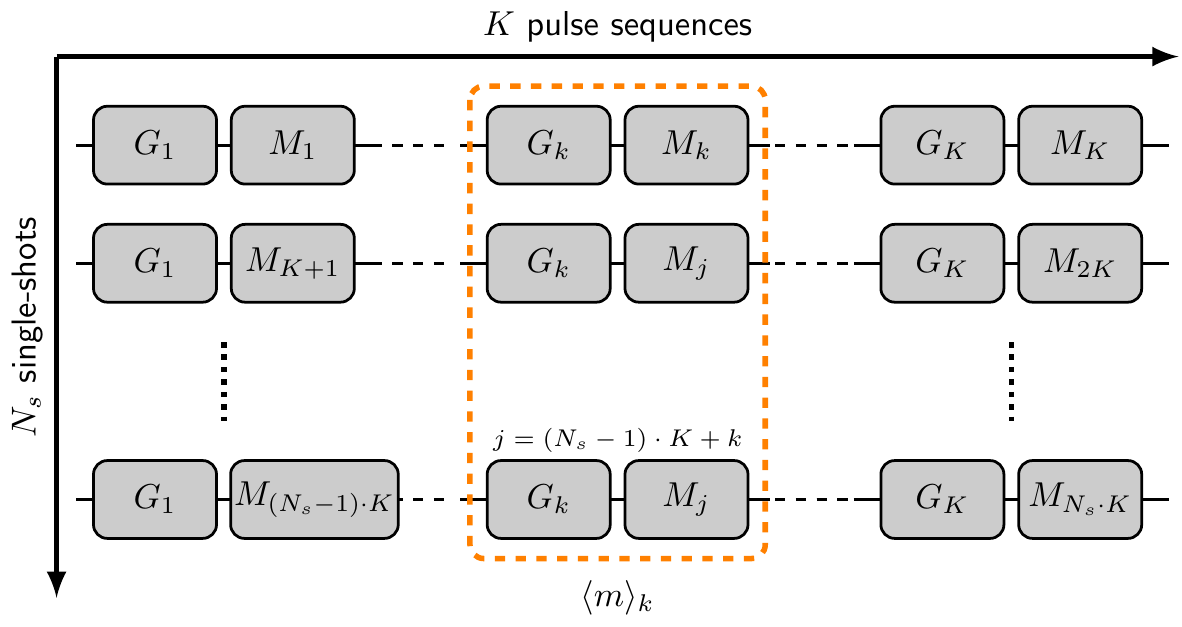}
			\caption{\label{fig:structure}
				Illustration of the order in which measurements are done.
				We measure the effect of $K$ different quantum operations $G_k$ where $k=1,..., K$ with a projective measurement $M_k$ done immediately after each $G_k$.
				All $K$ quantum operations are measured consecutively, and the experiment is repeated $N_s$ times.
				When the qubit is reset in between measurements, all $N_s$ single-shot measurements related to $G_k$ can be averaged to create the average IQ-point $\langle m\rangle_k$.
			}
		\end{center}
	\end{figure}

	The system consists of two transmon-type fixed-frequency superconducting qubits \cite{Koch2007a} coupled by a flux tunable coupler \cite{McKay2016, Roth2017}.
	Experiments are carried out on one of the qubits with a transition frequency of $\omega_{01}/2\pi = 5117.22~\rm{MHz}$, an anharmonicity $\Delta/2\pi = -315.28~\rm{MHz}$ and coherence times of  $50~\mu\rm{s}$ and $39~\mu\rm{s}$ for $T_1$ and $T_2$, respectively.
	
	In our setup we measure the effect of pulse sequences $k=1,...,K$, each viewed as an operation $G_k$, with a projective measurement $M_k$ of the qubit.
	This measurement is implemented by probing a dispersively coupled readout resonator with a square pulse modulated at its resonance frequency of $6.841~\rm{GHz}$ \cite{Blais2004}.
	The reflected signal goes through a travelling wave parametric amplifier \cite{Macklin2015, Roy2015} which allows us to perform the readout with a $2.5~\mu\rm{s}$ long pulse.
	After further amplification the signal is down-converted to $\omega_\text{IF}/2\pi=10~\rm{MHz}$ and digitized by an ADC resulting in the signal $s(t)={\rm Re}[Se^{i\theta}e^{i\omega_{IF}t}]$, where $S$ and $\theta$ depend on the qubit state \cite{Krantz2019,Bianchetti2009}.
	We represent $S$ and $\theta$ in the IQ-plane as an IQ-point $m=(I, Q)$ given by $I+iQ=Se^{i\theta}$.
	
	We first measure $G_1$ to $G_K$ and repeat the measurement $N_s$ times to gather $N_s$ single-shots for each $G_k$ \cite{Mallet2009}.
	To keep track of time ordering we assign an index $j$ to each measurement $M_j$ and its outcome $m_j$.
	For example, the $i$-th measurement of operation $G_k$ corresponds to measurement $M_{k+iK}$, see Fig.~\ref{fig:structure}.
	
	\begin{figure}
		\centering
		\includegraphics[width=\columnwidth]{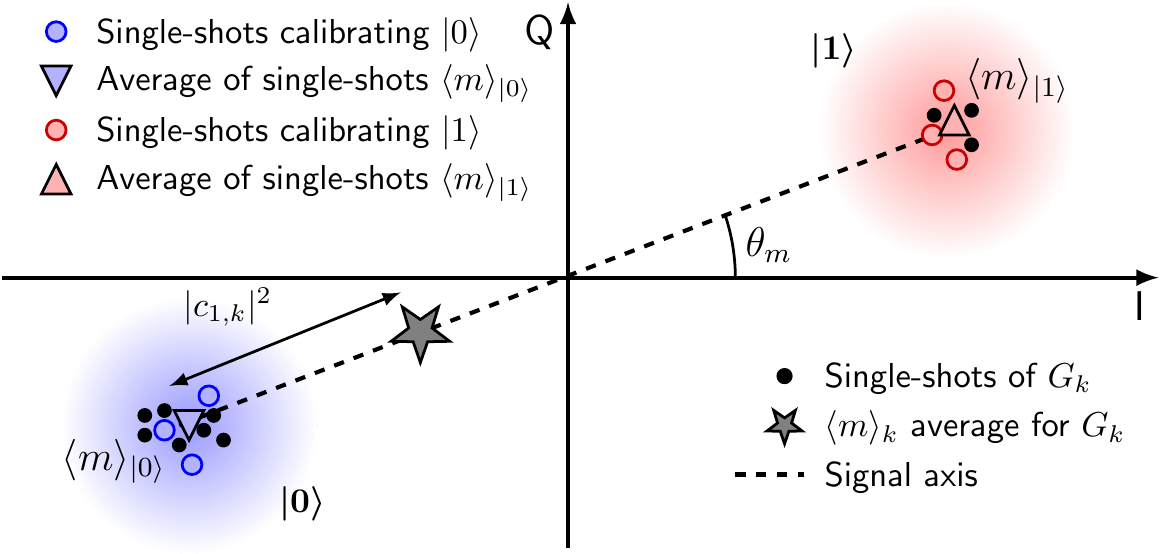}
		\caption{
			Illustration of the signal axis with angle $\theta_m$ defined by $\langle m\rangle_{\ket{0}}$ and $\langle m\rangle_{\ket{1}}$.
			The IQ-points form two clusters corresponding to $\ket{0}$ and $\ket{1}$.
			A pulse sequence $G_k$ that prepares the qubit in a superposition state results in single-shot IQ-points (black dots) distributed between the $\ket{0}$ and $\ket{1}$ states.
			The average of these single-shots (gray star) lies on the signal axis.}
		\label{fig:axis_sketch}
	\end{figure}
	
	When the qubit is initially in the ground state $\ket{0}$, $G_k$ creates a state $\ket{\Psi_k} = c_{0,k}\ket{0}+c_{1,k}\ket{1}$ with $|c_{0,k}|^2 + |c_{1,k}|^2 =1$.
	The measurement projects $\ket{\Psi_k}$ onto the ground state or the excited state $\ket{1}$ of the qubit with probabilities $|c_{0,k}|^2$ and $|c_{1,k}|^2$, respectively \cite{Blais2004}.
	Therefore, single-shot measurements form two clusters of points in the IQ-plane corresponding to $\ket{0}$ and $\ket{1}$.
	By averaging the $N_s$ single-shot measurements of $\ket{\Psi_k}$ we obtain an IQ-point
	\begin{align}\label{eqn:axis}
	\langle m\rangle_k =\langle m\rangle_{\Psi_k}= \langle m\rangle_{\ket{0}} + |c_{1,k}|^2\left(\langle m\rangle_{\ket{1}}-\langle m\rangle_{\ket{0}}\right),
	\end{align}
	where $\langle m\rangle_{\ket{i}}$ is the average IQ-point of state $\ket{i}$.
	$\langle m\rangle_k$ is therefore located on the \emph{signal axis} between the two IQ-points $\langle m\rangle_{\ket{0}}$ and $\langle m\rangle_{\ket{1}}$.
	The orientation of this axis is defined by an angle $\theta_m$ to the in-phase axis $I$, see Fig.~\ref{fig:axis_sketch}.
	The distance on this axis between $\langle m\rangle_{\ket{0}}$ and $\langle m\rangle_k$ forms the signal of $G_k$ interpreted as the probability that the qubit is in state $\ket{1}$ after $G_k$ is applied.
	The signal axis is typically found by a singular value decomposition (SVD) applied to the average IQ-points $\langle m\rangle_k$.
	
	\section{Restless measurements\label{sec:restless_intro}}
	
	\begin{figure}[!t]
		\begin{center}
			\includegraphics[width=\columnwidth]{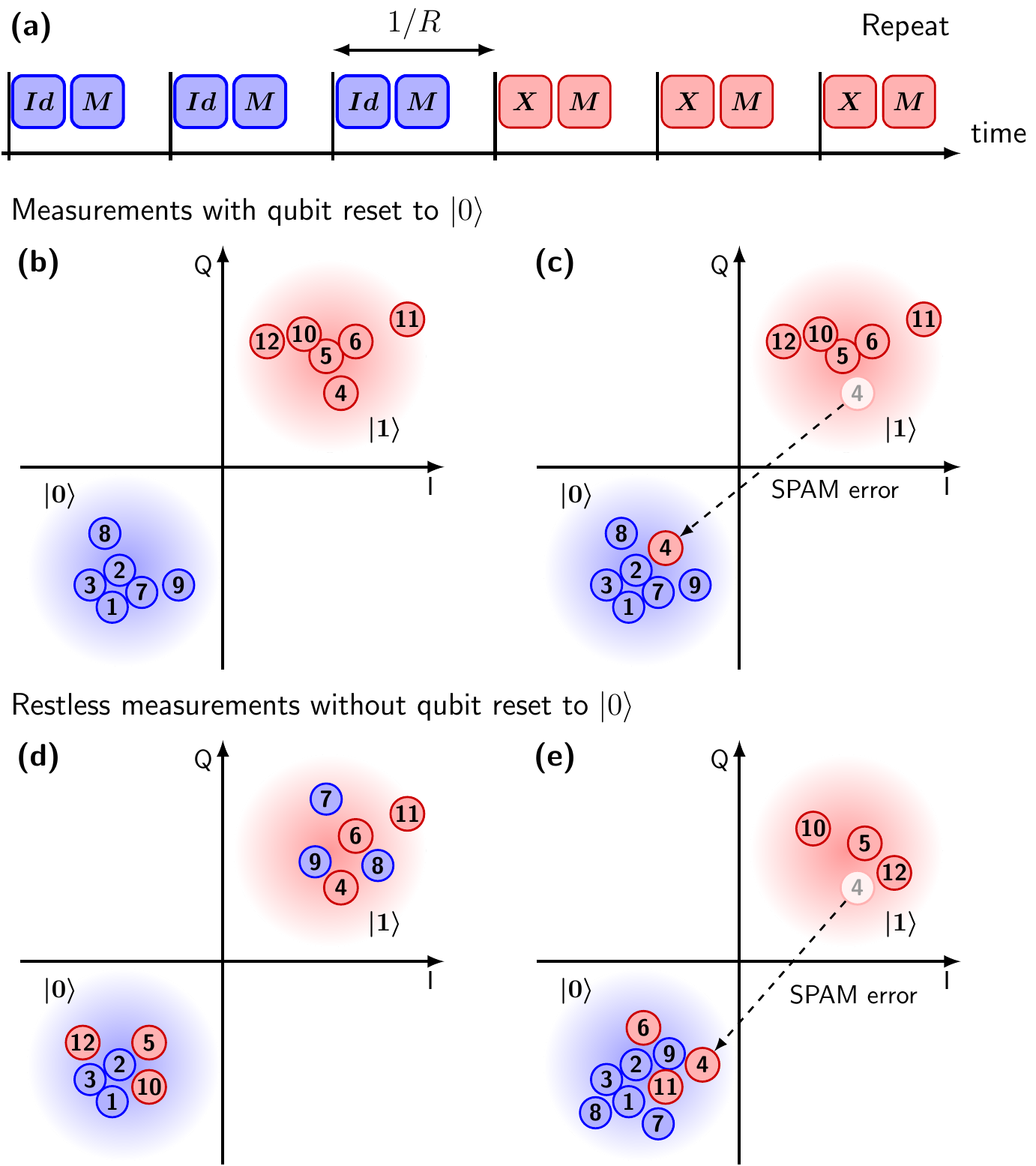}
			\caption{\label{fig:concept}
				Illustration of the difference between standard measurements with $T_1$ reset (b), (c) and restless (d), (e).
				The pulse sequence with $2N=6$, shown in (a), is repeated twice. 
				In (b)-(e) the numbers in the IQ-points indicate the order in which they were acquired.
				The ideal outcome for standard measurements (b) is minimally affected by a SPAM error occurring at the fourth measurement (c).
				The ideal IQ-points of restless measurements (d) are strongly affected by a SPAM error (e) which modifies all subsequent measurement outcomes.
			}
		\end{center}
	\end{figure}
	
	We now illustrate the difference between restless measurements and standard measurements using a pulse sequence made of $2N$ gates $G_k$ each followed by a measurement $M$.
	The first $N$ gates are identity gates ($G_k=Id$) while the next $N$ gates are $\pi$-rotations about the $x$-axis of the qubit Bloch sphere ($G_k=X$).
	These $2N$ measurements are repeated $N_\text{s}$ times, see Fig.~\ref{fig:concept}(a).
	The pulses implementing the readout and control operation $M_j\circ G_k$ are applied immediately after each trigger pulse. 
	In the standard operation mode the trigger rate is $R=1~\rm{kHz}$ to reset the qubit to the ground state by $T_1$-decay.
	Due to this initialization, the pulse sequences $G_k= Id$ and $G_k= X$ prepare the ground and excited state of the qubit, respectively, see Fig.~\ref{fig:concept}(b).
	State preparation and measurement (SPAM) errors have a minimal impact on the single-shots, see Fig.~\ref{fig:concept}(c).
	
	To perform restless measurements we increase the trigger rate to $R=100~\rm{kHz}$. 
	The qubit is not reset to its ground state after each measurement.
	The initial state for a given sequence $M_j\circ G_k$ is thus the outcome of the previous measurement $M_{j-1}$.
	As a result, the measurement outcomes of the $Id$ and $X$ operations no longer match with the ground and excited state of the qubit, respectively, see Fig.~\ref{fig:concept}(d).
	Instead, they are distributed between the ground and excited states.
	Since a SPAM error (for instance, caused by gate errors, $T_1$-decay, or unwanted qubit excitations) may change the initial state, the distribution of IQ-points is randomized between the ground and excited state, see Fig.~\ref{fig:concept}(e).
	Any operation $G_k$ that creates a superposition state will further randomize restless IQ-points, as the measurement induced projection into the ground or the excited state is a probabilistic process. 
	These random processes are not an issue when the qubit is reset to the ground state in between measurements.
	
	\subsection{Restless signals\label{sec:restless_signal}}
	
	\begin{figure*}[htbp!]
		\begin{center}
			\includegraphics[width=\textwidth]{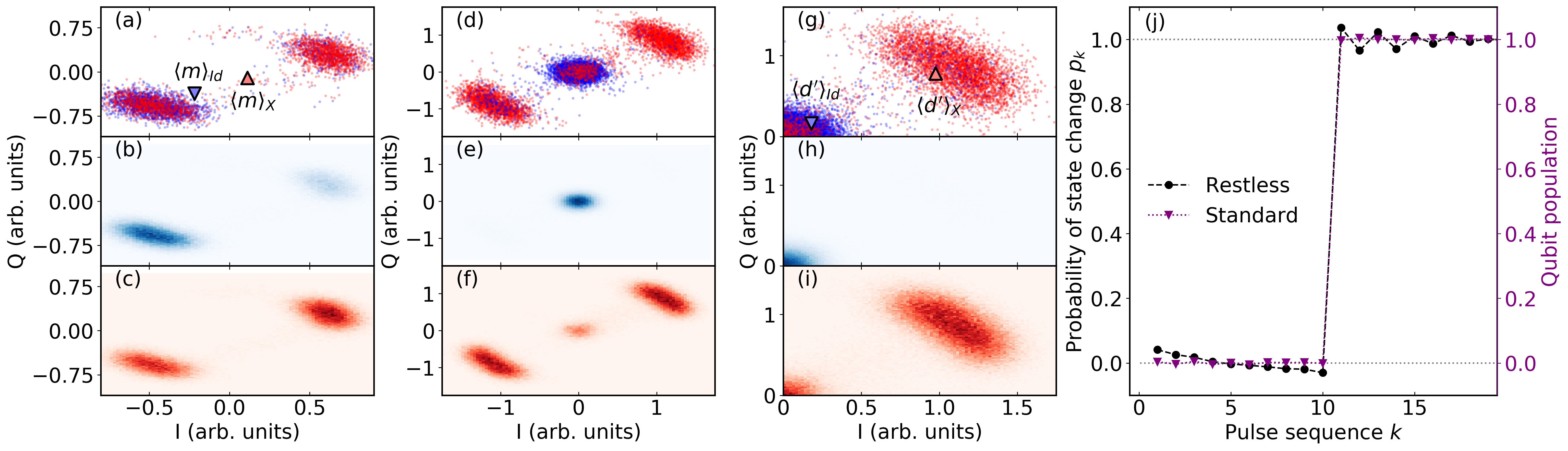}
			\caption{\label{fig:axis}
				Efficient reconstruction of the restless signal axis. 
				Here, $G_k=Id$ for $k=1,...,10$ (blue data) and $G_k=X$ for $k=11,...,20$ (red data).
				$20\times N_\text{s}=2\cdot 10^5$ single-shot IQ-points are measured at a $100~\rm{kHz}$ repetition rate.
				(a), (d), and (g) show the measured single-shots $m_j$, the difference IQ-points $d_j$ and the folded IQ-points $d'_j$, respectively. 
				Only the first 2.5\% of that data are shown to avoid overcrowding the figure. 
				(b), (e), and (h) show all the single-shot data for which $G_k=Id$ represented as two-dimensional density plots for $m_j$, $d_j$, and $d'_j$, respectively.
				(c), (f), and (i) show the data for $G_k=X$.
				The triangles in (a) show the average IQ-points $\langle m\rangle_{Id}$ and $\langle m\rangle_X$ and the triangles in (g) show the average IQ-points $\langle d'\rangle_{Id}$ and $\langle d'\rangle_X$.
				(j) Restless signal (black dots) obtained from Eq.~(\ref{eqn:restless_d_avg}) and rescaled to $[0,1]$ by estimating $\langle d' \rangle_{Id}$ and $\langle d' \rangle_{X}$ by averaging over $k=1,...,10$ and $k=11,...,20$, respectively.
				The purple triangles show data acquired with a $1~\rm{kHz}$ repetition rate.
				Comparing both data sets reveals a distortion in the restless data.
			}
		\end{center}
	\end{figure*}
	
	As the single-shot restless measurements of $G_k$ are randomized between the ground and excited states, see Fig.~\ref{fig:concept}(d), their average outcome $\langle m \rangle_k$ converges to the origin of the IQ-plane, see Fig.~\ref{fig:axis}(a-c).
	Therefore, restless $\langle m \rangle_k$ do not measure the probability that $G_k$ excites the qubit and thus cannot be used as a signal as in Eq.~(\ref{eqn:axis}) and Fig.~\ref{fig:axis_sketch}.
	Instead, the state of each time ordered single-shot measurement $m_{j}$ must be compared to the previous single-shot $m_{j-1}$.
	We therefore assign a label $y_{j}\in\{A,B\}$ to each single-shot $m_j$ with a discriminator to distinguish between ground and excited state.
	We can analyze restless data without knowing whether $A$ or $B$ corresponds to the qubit ground state.
	The restless signal is then defined as the average of an indicator function, i.e.
	\begin{equation}\label{eq:restless_signal}
	s_k=N_s^{-1}\sum_{i=0}^{N_s-1}\mathbf{1}_{k+iK},
	\end{equation}
	where $\mathbf{1}_{j}=1$ if $y_{j}\neq y_{j-1}$ and $\mathbf{1}_{j}=0$ otherwise \cite{Rol2017}.
	Therefore, $s_k$ is a measurement of the probability $p_k$ that $G_k$ \emph{changes} the qubit state.
	
\begin{figure}
		\centering
		\includegraphics[width=\columnwidth]{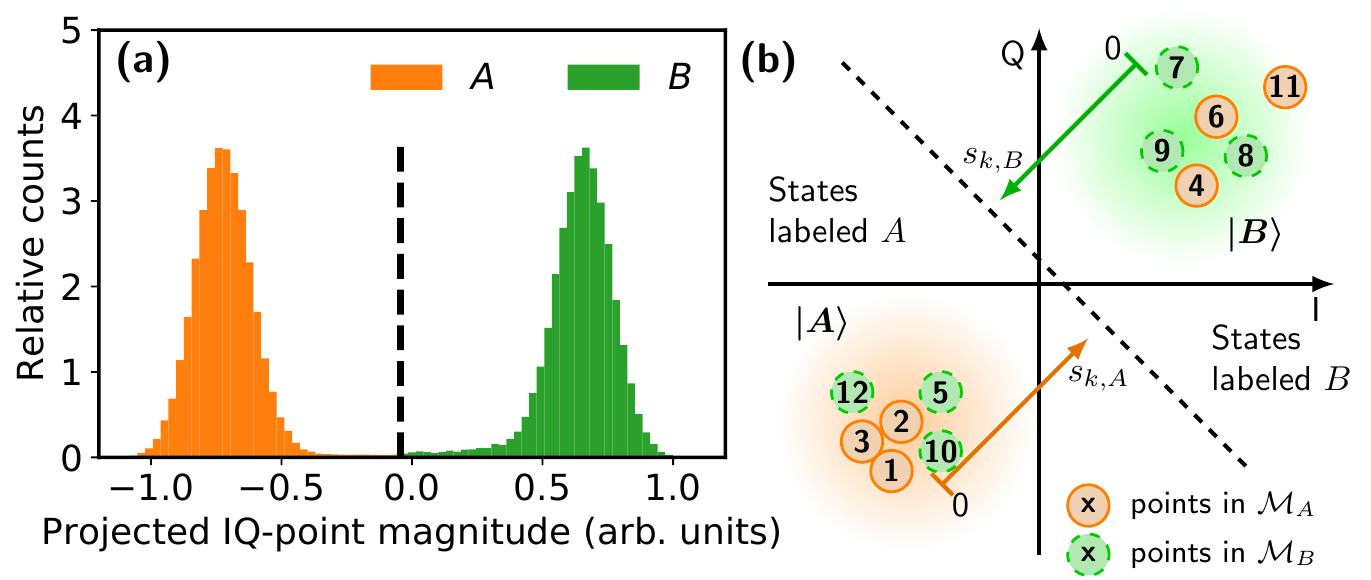}
		\caption{(a) One-dimensional discriminator built by projecting the $m_j$ IQ-points onto the signal axis. The discriminating line is the average of the 1\% and 99\% quantile. Since we cannot determine which peak is the ground or excited state we assign them the labels $A$ and $B$.
			(b) We recolored Fig.~\ref{fig:concept}(d) to illustrate which IQ-points belong to $\mathcal{M}_A$ and $\mathcal{M}_B$ and indicate the signals $s_{k,A}$ and $s_{k,B}$ obtained from Eq.~(\ref{eqn:ska_skb}) as arrows. The dashed line is a discriminator.}
		\label{fig:discriminator}
	\end{figure}

	To assign the labels $A$ and $B$ we need a discriminator.
	To avoid the computational cost of training a clustering algorithm on the two-dimensional single-shot restless IQ-points $m_j$, see Appendix~\ref{sec:clustering_cost}, we build a discriminator by projecting each $m_j$ on the signal axis (shown in Fig.~\ref{fig:axis_sketch}) which creates a one-dimensional bi-modal distribution.
	The discriminating line is obtained as the average of the 1\% and 99\% quantiles, see Fig.~\ref{fig:discriminator}(a). 
	
	To build our discriminator we require the signal axis in the IQ-plane which maximizes the separation of the two states.
	Since the averages $\langle m\rangle_k$ converge to the origin of the IQ-plane they cannot reliably be used as input to a SVD as done in the conventional, non-restless case, see Sec.~\ref{sec:setup}.
	Indeed, estimating the angle of a signal axis going through noisy points becomes less reliable the closer the points are to each other.
	
	\subsection{Efficient reconstruction of restless signals\label{sec:restless_axis}}
	
	We now show how to efficiently recover the signal axis from the single-shot restless data which we illustrate with the same pulse sequences as described in Sec.~\ref{sec:restless_intro} with $N=10$.
	First, we subtract from each measured IQ-point $m_j$ the point $m_{j-1}$ obtained in the previous measurement.
	The resulting difference points $d_j=m_j-m_{j-1}$ form three clusters in the IQ-plane see Fig.~\ref{fig:axis}(d).
	The cluster centered around $(0,0)$ corresponds to all the measured outcomes for which $M_j\circ G_k$ did not change the qubit state, i.e. $y_{j}=y_{j-1}$. 
	The two clusters that are not centered around $(0, 0)$ are measured outcomes for which $M_j\circ G_k$ changed the qubit state, i.e. $y_{j}\neq y_{j-1}$, see Fig.~\ref{fig:axis}(e) and (f).
	Next, we compute the IQ-points $d'_j=|{\rm Re}(d_j)|+i|{\rm Im}(d_j)|$ to project the data into the first quadrant of the IQ-plane, therefore combining all $M_j\circ G_k$ that changed the qubit state into a single cluster.
	Since the $d'_j$ measure a \emph{state change} of the qubit (instead of the state as measured by $m_j$), the average
	\begin{align}
	\langle d'\rangle_{k}=N_s^{-1}\sum_{i=0}^{N_s-1}d'_{k+iK}
	\end{align}
	does not converge to the origin of the IQ-plane as does $\langle m\rangle_k$.
	We may apply a SVD to the averaged points $\langle d'\rangle_{k}$ to recover an axis with an angle $\theta_d$ in the IQ-plane which yields the highest signal to noise ratio (SNR).
	The signal axis needed for the discriminator processing the single-shots $m_j$ has an angle $\theta_m$, as in Fig.~\ref{fig:axis_sketch}, given by either $\theta_d$ or $\pi-\theta_d$ due to the folding $d_j\to d'_j$ into the first quadrant.
	In our analysis we try both angles and use the one with the highest SNR to build the discriminator which allows us to reconstruct the restless signal $s_k$.
	
	We observe that $\langle d'\rangle_k$ is an affine transformation of $s_k$ since
	\begin{align} \label{eqn:restless_d_avg}
	\langle d'\rangle_k=\langle d'\rangle_{Id} + p_k\big(\langle d'\rangle_{X}-\langle d'\rangle_{Id}\big).
	\end{align}
	Therefore, $\langle d' \rangle_k$ is a measurement of the state change probability $p_k$ and is an alternative to the restless signal $s_k$.
	The end points of this axis $\langle d'\rangle_{Id}$ and $\langle d'\rangle_{X}$ correspond to the $Id$ and $X$ operators.
	They can be measured with calibration sequences to obtain a restless signal normalized to the interval $[0, 1]$ as in Eq.~(\ref{eq:restless_signal}), see Fig.~\ref{fig:axis}(j).
	
	\subsection{Correcting distortions\label{sec:bias}}
	
	We compare the restless signal to a conventional measurement of the same pulse sequences, i.e. done at a $1~\rm{kHz}$ repetition rate to initialize the qubit to $\ket{0}$, see Fig.~\ref{fig:axis}(j). 
	The restless signal shows a distortion in which identical pulse sequences produce different results.
	The first ten measurements, for which $G_k=Id$, exhibit an exponential decay and the last ten measurements for which $G_k=X$ have a zigzag pattern.
	
	At the $R=100~\rm{kHz}$ repetition rate the idle time between the end of the measurement pulse and the next pulse sequence is $\sim7.5~\mu\rm{s}$.
	Measurements $m_j$ that yield the excited state will experience $T_1$-decay during this  $7.5~\mu\rm{s}$ and will thus result in a higher probability of a SPAM error in the subsequent outcome $m_{j+1}$.
	Therefore, to remove this distortion we analyze the restless data using a discriminator, as discussed in Sec.~\ref{sec:restless_axis}, and post-select each single-shot according to the previous single-shot to create two sets $\mathcal{M}_{A}=\{m_j\,|\,y_{j-1}= A\}$ and $\mathcal{M}_{B}=\{m_j\,|\,y_{j-1}= B\}$.
	Here, $\mathcal{M}_{A}$ and $\mathcal{M}_{B}$ collect the IQ-points where the previous outcome gave $A$ or $B$, respectively, see Fig.~\ref{fig:discriminator}(b).
	Since we use pulse sequences with $Id$ and $X$ operations we know if $m_j$ was obtained with $G_k=Id$ or $G_k=X$.
	This allows us to calculate the readout fidelities \cite{Magesan2015}
	\begin{align} \label{eqn:fa}
	\mathcal{F}_A =1-\frac{1}{2}\left[P_A(B|Id)+P_A(A|X)\right]=96.5\pm0.1\%
	\end{align}
	and 
	\begin{align}\label{eqn:fb}
	\mathcal{F}_B =1-\frac{1}{2}\left[P_B(A|Id)+P_B(B|X)\right]=82.9\pm0.3\%
	\end{align}
	for $\mathcal{M}_A$ and $\mathcal{M}_B$, respectively, see Appendix~\ref{sec:SPAM}.
	The confidence intervals are obtained using Jeffreys interval at a 95\% confidence level \cite{Brown2001}.
	Here, $P_x(y|G)$ is the probability to measure $y$ given the operation $G$ and initial state $x$ and measures the SPAM error for the choices of $x$, $y$, and $G$ in Eqs.~(\ref{eqn:fa}) and (\ref{eqn:fb}).

	The different readout fidelities $\mathcal{F}_A$ and $\mathcal{F}_B$ together with the relative number of measurements in  $\mathcal{M}_A$ and $\mathcal{M}_B$ create the observed distortions in the restless signal.
	Due to the randomization discussed in Sec.~\ref{sec:restless_intro} and the $T_1$-decay the probability that the single-shot measurements associated to $G_k$ belong on average to $\mathcal{M}_A$, labeled by $p_{k, A}$, is a function of all previous operations $G_k$, see Fig.~\ref{fig:bias}(a) and the model in Appendix \ref{sec:restless_model}.
	The restless signal $s_k$ is thus a weighted average of the signals $s_{k,A}$ and $s_{k,B}$, shown in Fig.~\ref{fig:discriminator}(b), obtained from Eq.~(\ref{eq:restless_signal}) by restricting the calculation to the IQ-points in $\mathcal{M}_A$ and $\mathcal{M}_B$, respectively,~i.e.
	\begin{align}\label{eqn:ska_skb}
	s_k=&\,\frac{|\mathcal{M}_A|_k}{N_s|\mathcal{M}_A|_k}\sum_{\mathclap{\substack{i=0\\ m_{k+iK}\in\mathcal{M}_A}}}^{N_s-1}\mathbf{1}_{k+iK} + \frac{|\mathcal{M}_B|_k}{N_s|\mathcal{M}_B|_k}\sum_{\mathclap{\substack{i=0\\ m_{k+iK}\in\mathcal{M}_B}}}^{N_s-1}\mathbf{1}_{k+iK} \\
	=&\,p_{k,A}s_{k,A}+(1-p_{k,A})s_{k,B}\quad\text{with}\quad p_{k,A}=\frac{|\mathcal{M}_A|_k}{N_s}. \notag
	\end{align}
	Here, $|\mathcal{M}_{A,B}|_k$ indicates the number of measurements belonging to $G_k$ which are in $\mathcal{M}_{A,B}$.
	This weighted average is confirmed in Fig.~\ref{fig:bias}(b) which shows how the restless signal varies between $s_{k,A}$ and $s_{k,B}$ with a pattern that matches the weights $p_{k,A}$ in Fig.~\ref{fig:bias}(a).
	We attribute the lower signal range of $s_{k,B}$ compared to $s_{k,A}$, i.e $[13\%, 83\%]$ and $[2\%, 95\%]$, respectively, to SPAM errors, see Appendix~\ref{sec:SPAM}.
	More concretely, from the readout fidelities in Eq.~(\ref{eqn:fa}) and Eq.~(\ref{eqn:fb}) we can now identify $B$ as the excited state as $s_{k,B}$ has higher SPAM errors mainly due to $T_1$-decay.
	Both signals $s_{k,A}$ and $s_{k,B}$ can be analyzed either individually or calibrated to remove SPAM errors \cite{Bravyi2020} and combined into a single data set.
	Alternatively, retaining only the data set with the highest SNR reduces the number of single-shots analyzed but can benefit some measurements such as RB, see Sec.~\ref{sec:tune_up}.
	
	\begin{figure}[!t]
		\begin{center}
			\includegraphics[width=\columnwidth,clip, trim=0 6 0 6]{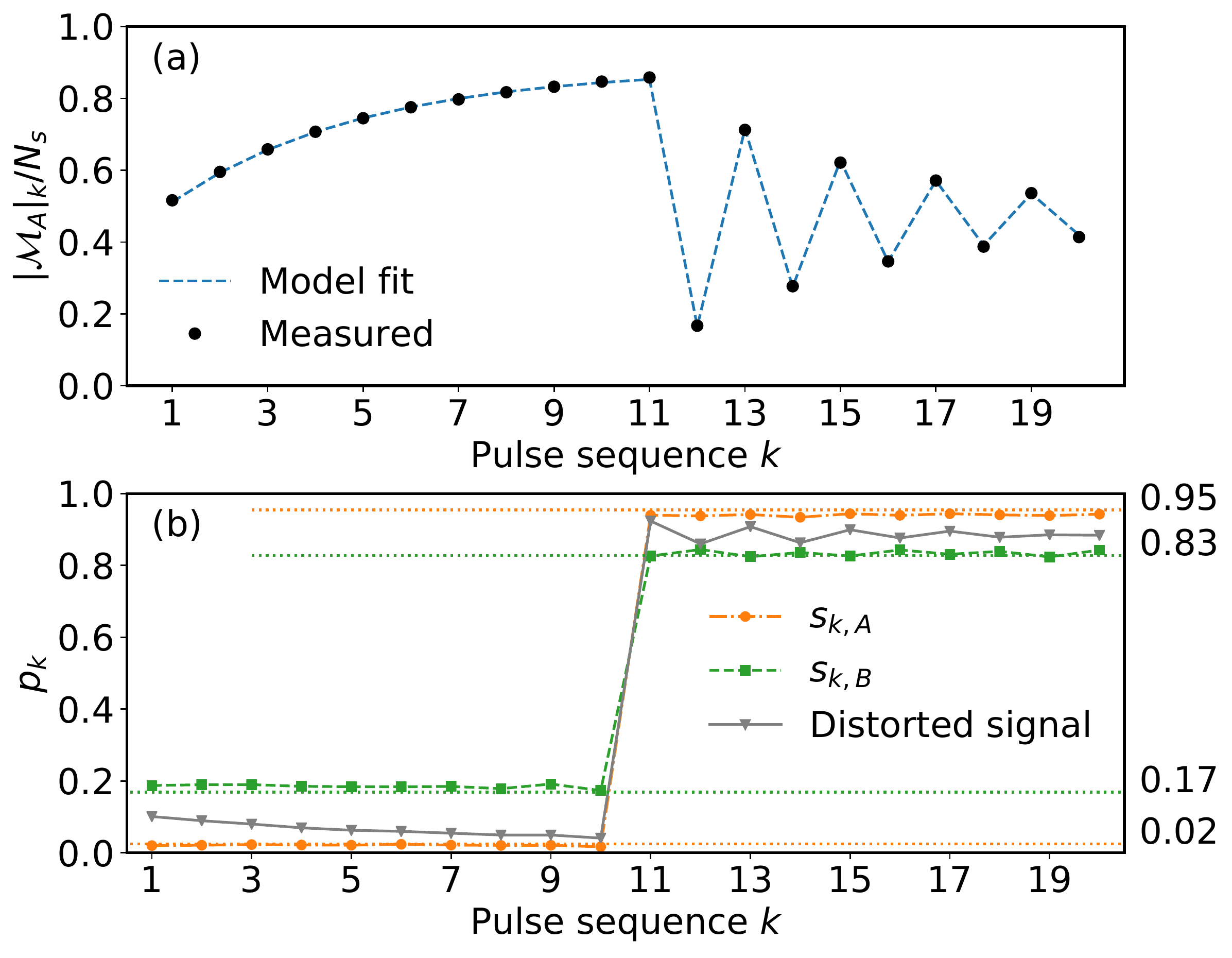}
			\caption{\label{fig:bias}
				(a) Probability that $m_j\in\mathcal{M}_A$, estimated by $|\mathcal{M}_A|_k/N_\text{s}$, as a function of the measurement index.
				The first and last ten pulse sequences correspond to $G_k=Id$ and $G_k=X$, respectively.
				The dashed line is a fit to the model discussed in Appendix \ref{sec:restless_model}.
				(b) Restless signals $s_{k,A}$ and $s_{k,B}$ built from $m_j\in\mathcal{M}_A$ and $m_j\in\mathcal{M}_B$, respectively.
				The dashed lines correspond to the measured SPAM errors shown in Fig.~\ref{fig:meas_histo}.
			}
		\end{center}
	\end{figure}

	\section{Restless calibration and characterization}
	
	Characterizing quantum chips and gates requires a broad variety of measurements in which the measured behavior is fit to a function describing the expected outcome.
	We anticipate that restless data acquisition can speed-up experiments such as spectroscopy, Ramsey and error amplifying gate sequences \cite{Sheldon2016} for both single- and two-qubit gates. 
	This frees up the hardware to run more quantum circuits.
	The signal of these experiments is often calibrated using the pulses sequences discussed in the previous sections.
	To illustrate calibration and characterization we measure a Rabi oscillation and perform RB of a single- and a two-qubit gate using restless measurements.
	
	\subsection{Rabi measurements\label{sec:rabi}}
	
	To measure a Rabi oscillation between the qubit states we apply a resonant Gaussian pulse with an amplitude $\alpha$ and subsequently measure the qubit.
	This pulse sequence is repeated for $K=128$ different, linearly-spaced, values $\alpha_k$ starting at $-90\%$ and ending at $90\%$ of the maximum arbitrary waveform generator output voltage of $0.8~\rm{V}$.
	Adding three $M\circ Id$ and three $M\circ X$ sequences allows us to mitigate readout errors by normalizing the measured signals to $[0,1]$. 
	
	\begin{figure}[!t]
		\begin{center}
			\includegraphics[width=\columnwidth,clip, trim=0 6 0 6]{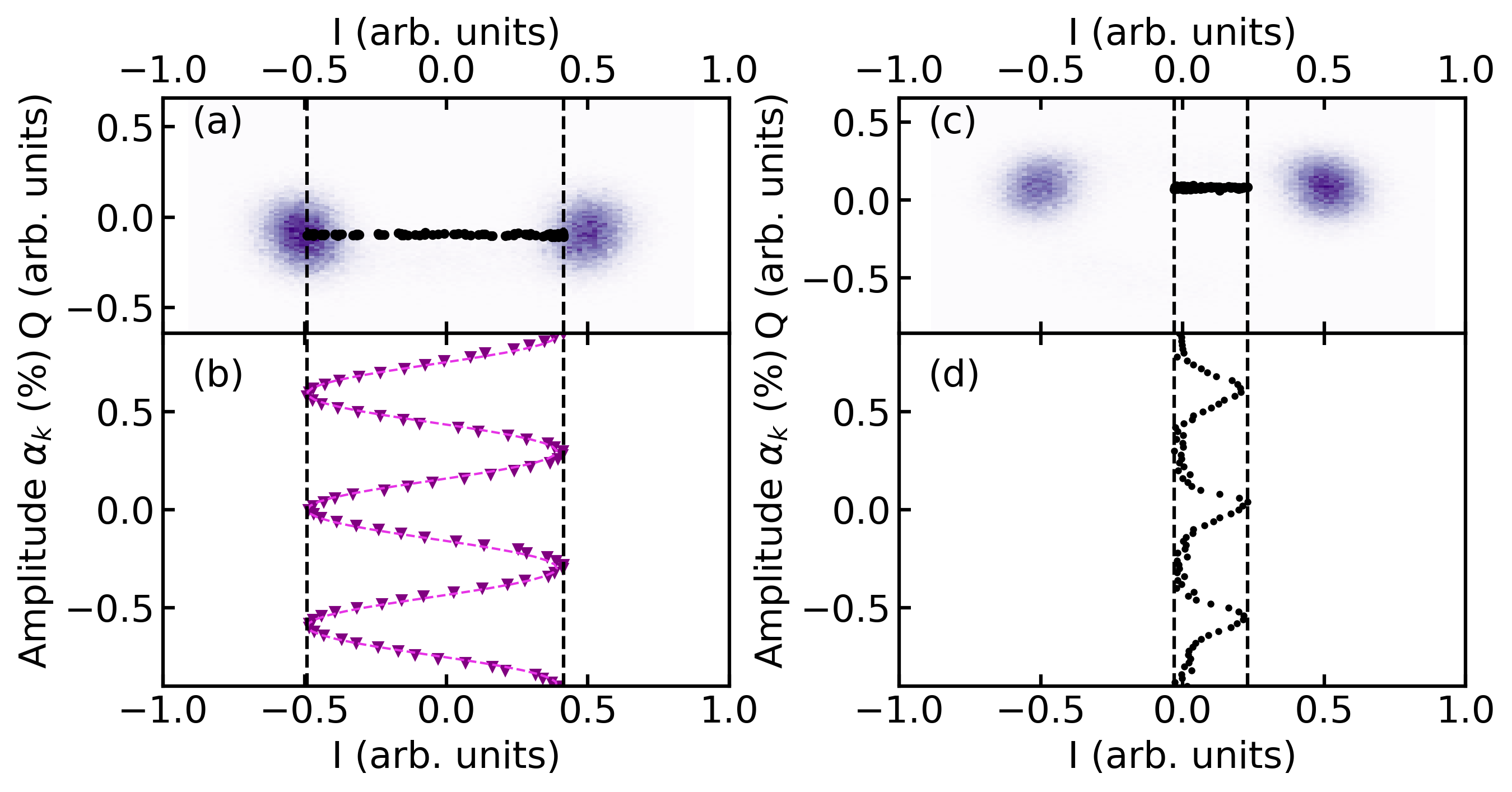}
			\caption{\label{fig:rabi}
				Density of IQ-points corresponding to a Rabi oscillation measured with a repetition rate of $1~{\rm kHz}$ (a) and $250~{\rm kHz}$ (c).
				A SVD applied to the average IQ points $\langle m\rangle_k$ yields the angle of the measurement axis $\theta_m$.
				The data are then rotated by this angle to align with the real axis for plotting.
				The average IQ-points $\langle m\rangle_k$ produce the expected Rabi oscillation for standard measurements (which we fit to a cosine), see (b), but unsurprisingly fail to reproduce the Rabi oscillation when measured at $250~\rm{kHz}$, see (d).
			}
		\end{center}
	\end{figure}
	
\begin{figure}[!t]
		\begin{center}
			\includegraphics[width=\columnwidth,clip, trim=0 6 0 6]{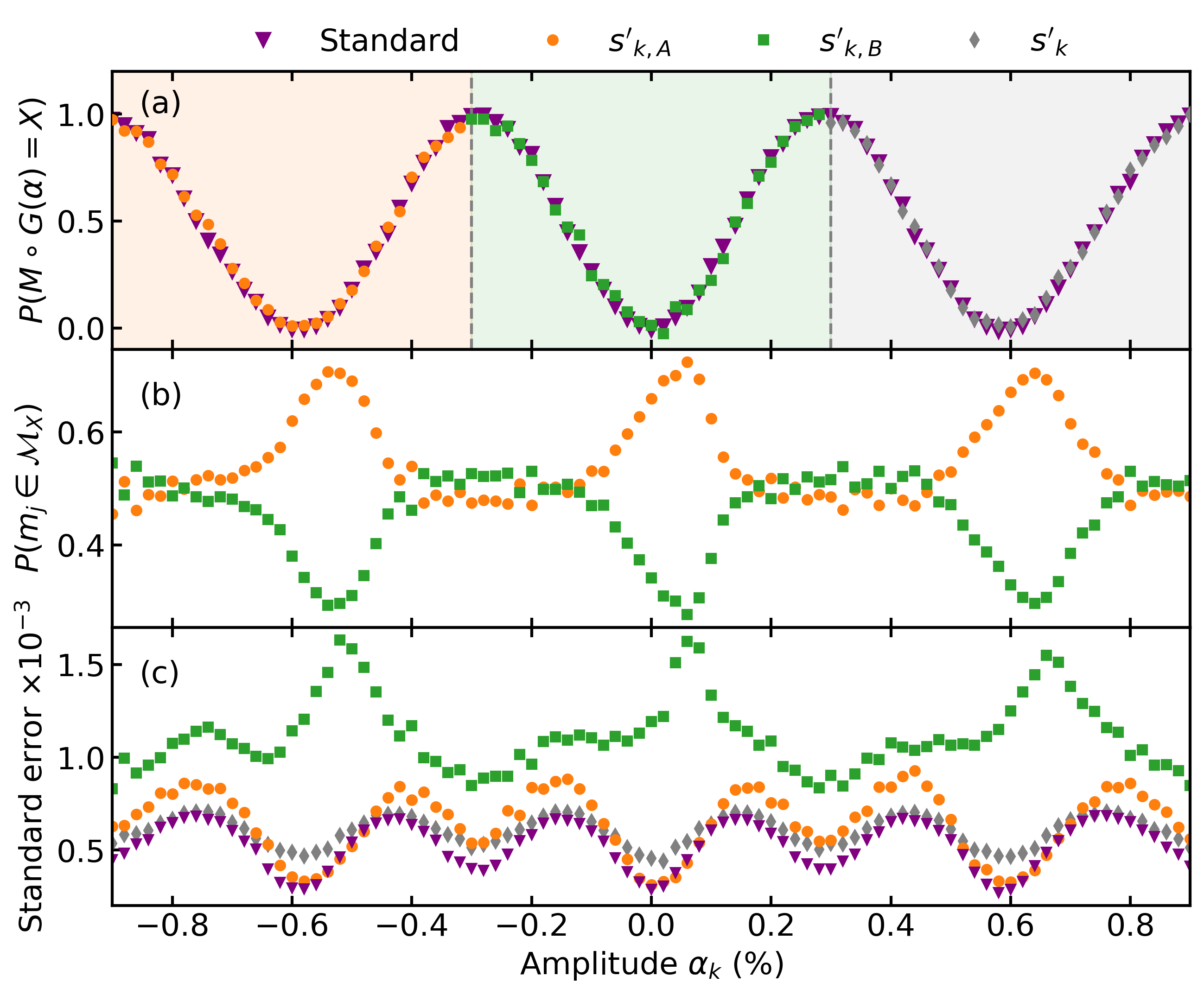}
			\caption{\label{fig:rabi_sig}
				(a) The restless-measured Rabi oscillation compared to the Rabi oscillation measured with a $1~\rm{kHz}$ repetition rate.
				For clarity, we only plot $s_{k,A}$ for $\alpha_k\in{-0.9,-0.3}$, $s_{k,B}$ for $\alpha_k\in{-0.3,0.3}$, and $s_{k}$ for $\alpha_k\in{0.3,0.9}$.
				(b) Measured probability that a restless IQ-point belongs to $\mathcal{M}_A$ or $\mathcal{M}_B$.
				(c) Standard error of the standard and restless signals.
				Since $\mathcal{M}_A$ has a lower standard error we conclude that $A$ is the ground state due to the asymmetric nature of $T_1$-decay.
			}
		\end{center}
	\end{figure}

	We measure a Rabi oscillation by gathering data at a $1~\rm{kHz}$ repetition rate, averaging the IQ-points $m_j$ over the $N_s=1000$ single-shots, and applying a SVD to find $\theta_m$, see Sec.~\ref{sec:setup}.
	This procedure nicely reveals the Rabi oscillation, see Fig.~\ref{fig:rabi}(a) and (b).
	Applying the same averaging procedure to analyze restless data, gathered at a trigger rate of $250~\rm{kHz}$, fails to reveal the Rabi oscillation, see Fig.~\ref{fig:rabi}(c) and (d).
	However, when we use the procedure outlined in Sec.~\ref{sec:restless_intro} to build the normalized readout-error mitigated restless signals $s_{k,A}$ and $s_{k,B}$ we recover the Rabi oscillation without any distortions and in excellent agreement with the measurement at a $1~\rm{kHz}$ repetition rate, see Fig.~\ref{fig:rabi_sig}(a).
	We combine both signals as a weighted average 
	\begin{equation} \label{eqn:s_prime_k}
	s_k=(|\mathcal{M}_A|_ks_{k,A}+|\mathcal{M}_B|_ks_{k,B})/N_s
	\end{equation}
	which also shows the Rabi oscillation.
	We recover the amplitude response of the Rabi oscillation with a fit to a cosine function of the standard measurement, $s_{k,A}$, $s_{k,B}$, and $s_k$, yielding $0.5858(10)$, $0.5854(9)$, $0.5845(16)$, and $0.5853(10)~\rm{rad/V}$, respectively.
	All restless measurements reproduce the result of the standard measurement with $s_{k,B}$ slightly outside the one standard deviation interval.
	The confidence intervals reflect the standard error in the data sets shown in Fig.~\ref{fig:rabi_sig}(c) and shows that restless measurements have the same precision and accuracy as standard measurements.
	We anticipate that the methodology used here may be applied to other experiments such as Ramsey and gate error amplifying sequences used for calibration \cite{Sheldon2016}.

	\subsection{Randomized Benchmarking\label{sec:tune_up}}
	\begin{figure}[!t]
		\begin{center}
			\includegraphics[width=\columnwidth,clip, trim=0 6 0 6]{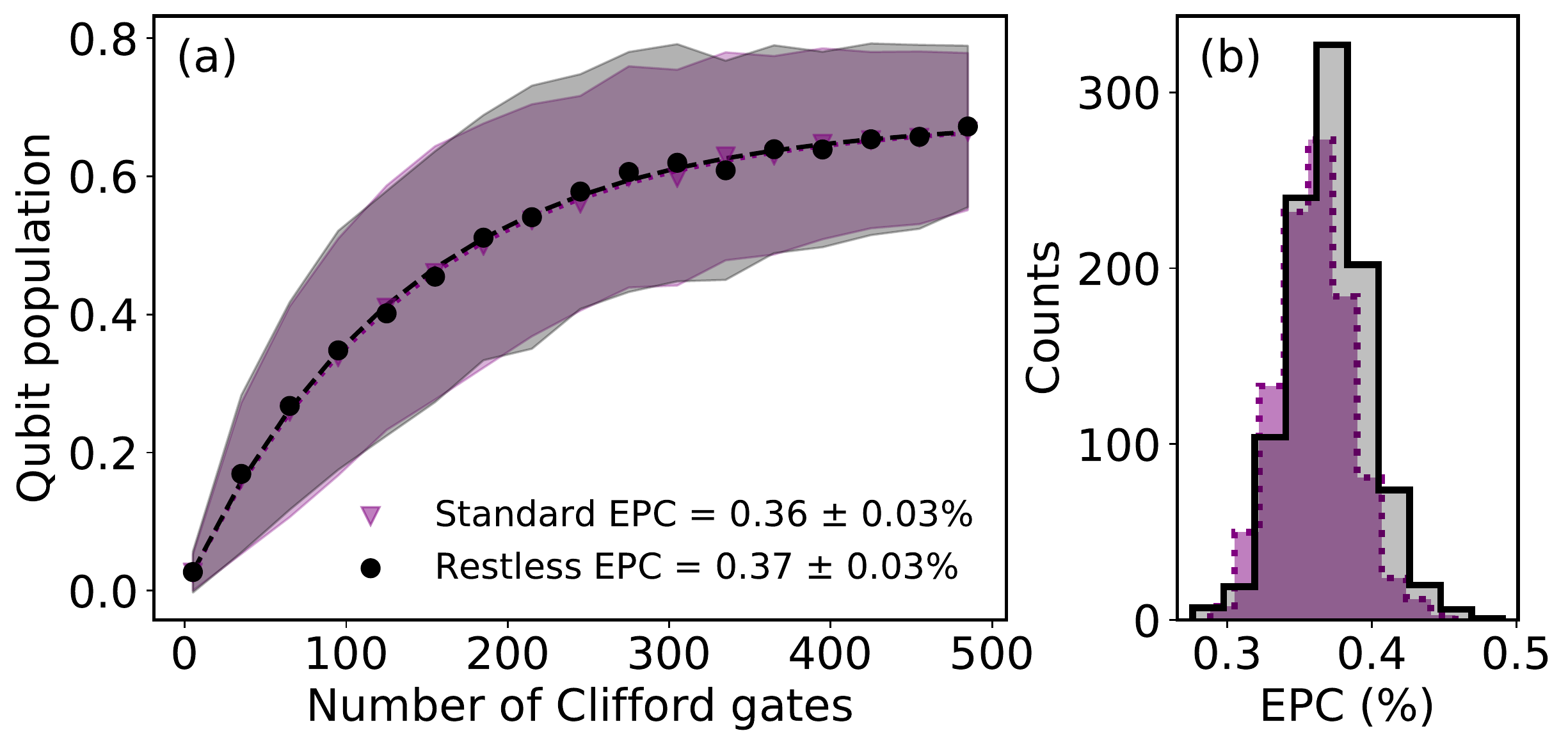}
			\caption{\label{fig:restless_RB}
				Randomized Benchmarking of single-qubit gates measured with a $1~\rm{kHz}$ and a $50~{\rm kHz}$ repetition rate.
				(a) The dot and shaded areas show the average and standard deviation, respectively, of all 200 realizations of random Clifford sequences.
				The averages are fit to the function $A + (B - A)\alpha^{N_c/2}$ with absolute weights given by the standard error of the mean. $A$, $B$, and $\alpha$ are fit parameter. The measured EPC is $(1-\alpha)/2$.
				(b) Distribution of 1000 measurements of the EPC. 
				Each measurement is obtained by randomly selecting 100 of the 200 sequences and fitting the mean as in (a). 
				The mean and standard deviation of the distribution in (b) are in the legend in (a).
			}
		\end{center}
	\end{figure}
	
	Randomized Benchmarking requires a substantial amount of measurements \cite{Ganzhorn2020}.
	Restless measurements increase the rate at which RB data is gathered.
	We characterize the single-qubit gates with 200 random sequences of $N_c$ Clifford gates, built from $X_{\pm\pi/2}$ and $Y_{\pm\pi/2}$ rotations, for 17 different values of $N_c$ ranging from 5 to 500.
	For each Clifford sequence we measure $N_s=2000$ single-shots at a $1~\rm{kHz}$ and at a $50~\rm{kHz}$ repetition rate.
	Since each single-qubit pulse lasts $4.16~{\rm ns}$ with a $4.16~{\rm ns}$ buffer in between pulses we limit the repetition rate of restless measurements to $50~{\rm kHz}$ to ensure that sequences with 500 Clifford gates do not overlap with the next measurement.
	Since the trigger rate is fixed, short Clifford sequences leave a long idle time before the start of the next sequence.
	$T_1$-induced errors of shorter Clifford sequences are thus magnified when the previous measurement produced a $\ket{1}$ state.
	To overcome this effect we post-select our data and use only the sequences initialized in $\ket{0}$.
	We therefore discard $60\%$ of the data and the effective restless rate is $20~{\rm kHz}$.
	To measure the EPC we randomly select 100 of the 200 Clifford sequences and fit their mean as shown in Fig.~\ref{fig:restless_RB}(a).
	This procedure is repeated 1000 times to create a distribution of the measured EPC, see Fig.~\ref{fig:restless_RB}(b).
	We measure an EPC of $0.36\pm0.03\%$ and $0.37\pm0.03\%$ for standard and restless measurements, respectively, where the error is the standard deviation of the distribution in Fig.~\ref{fig:restless_RB}(b).
	From a $z$-test with $z$-value $z=(0.36-0.37)/\sqrt{0.03^2+0.03^2}=-0.25$ we conclude that both methods yield an identical EPC with an $80\%$ confidence even after discarding 60\% of the restless data.
	
	Gathering $6.8$ million single-shots takes $113.3~{\rm minutes}$ and $2.3~{\rm minutes}$ at $1~\rm{kHz}$ and $50~\rm{kHz}$, respectively, which emphasizes the gain of high repetition rates.
	These figures do not include the constant $3$ minutes needed to prepare the gate sequences, initialize the hardware, and transfer the pulse data to the arbitrary waveform generators.
	
	\begin{figure}
		\centering
		\includegraphics[width=\columnwidth,clip, trim=0 6 0 6]{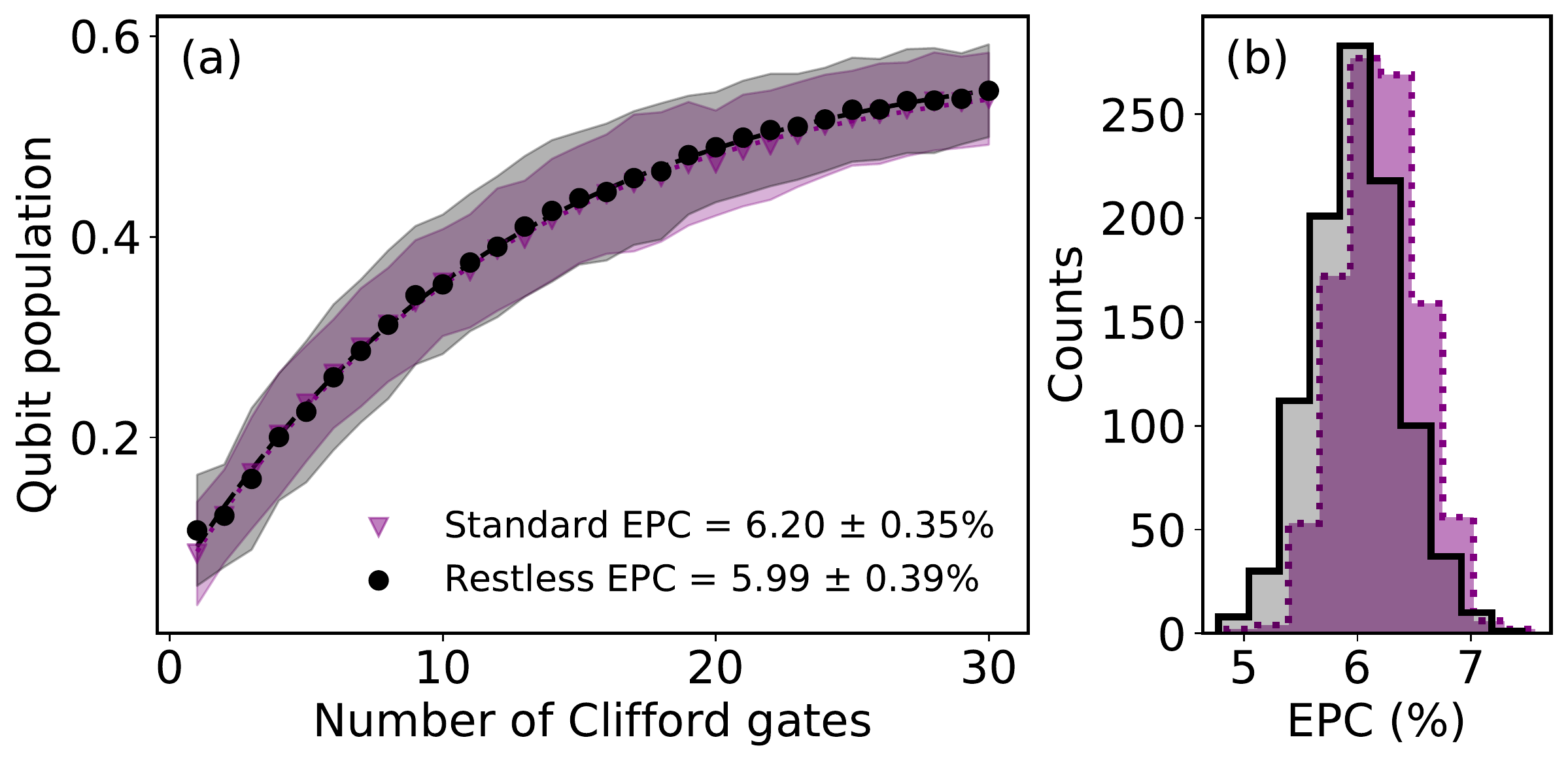}
		\caption{
			Randomized Benchmarking of a roughly calibrated two-qubit CZ gate measured with a $1~\rm{kHz}$ and a $20~{\rm kHz}$ repetition rate.
			The dot and shaded areas show the average and standard deviation, respectively, of 90 realizations of random Clifford sequences.
			The averages are fit to the function $A + (B - A)\alpha^{N_c/2}$ with absolute weights given by the standard error of the mean. The measured EPC is $(1-\alpha)3/4$.
			(b) Distribution of 1000 measurements of the EPC. 
			Each measurement is obtained by randomly selecting 50 of the 90 sequences and fitting the mean as in (a). 
			The mean and standard deviation of the distribution in (b) are in the legend in (a).
		}
		\label{fig:cz_rb}
	\end{figure}
	
	In addition to single-qubit RB we measure the EPC of a roughly calibrated two-qubit CZ gate, calibrated on a different chip with an identical architecture. 
	The CZ gate is implemented by modulating the tunable coupler with an oscillating magnetic flux with a frequency $\omega_\Phi=\omega_{10}-\omega_{20}$ which drives the transition between the $\ket{20}$ and $\ket{11}$ states \cite{Strauch2003, Ganzhorn2020, Bengtsson2019}.
	We characterize the CZ gate by measuring 90 random sequences of Clifford gates with up to 30 Clifford gates.
	Since each CZ gate lasts $333~{\rm ns}$ we limit the repetition rate of restless measurements to $20~{\rm kHz}$ to ensure that sequences with 30 Clifford gates do not overlap with the next measurement.
	As before, we post-select the restless data keeping only measurements initialized in $\ket{00}$ to mitigate $T_1$-induced errors.
	We therefore discard $60\%$ of the data and the effective restless rate is $8~{\rm kHz}$.
	As before, we bootstrap the measurement by determining the EPC 1000 times where we randomly select 50 of the 90 sequences.
	At $1~{\rm kHz}$ and $20~{\rm kHz}$ we measure an EPC of $6.20\pm0.35\%$ and $5.99\pm0.39\%$, respectively, see Fig.~\ref{fig:cz_rb}.
	From a $z$-test with $z$-value $z=(6.20-5.99)/\sqrt{0.35^2+0.39^2}=0.40$ we conclude that both methods yield an identical EPC with an $69\%$ confidence.
	
	\section{Discussion and Conclusion \label{sec:conclusion}}
	
	We have demonstrated how to analyze single-shot data gathered at elevated repetition rates without qubit reset.
	We show how to efficiently build a state discriminator and use it to correct distortions in the restless signal.
	Our work extends the method of Rol \emph{et al.} \cite{Rol2017} to common qubit calibration and characterization experiments.
	Our approach post-selects the single-shots according to their initial state.
	The signals of each data set can be recombined once measurement biases that depend on the initial state are corrected.
	We argued that the initial state of the data set with the lowest variance is $\ket{0}$.
	In our experience retaining only this data set produces the highest SNR despite not using all the data.
	We illustrate our method by showing that a restless-measured Rabi oscillation -- as an exemplary measurement for calibration -- and a RB measurement reliably reproduce the results obtained at low trigger rates.
	
	Restless measurements can be applied without any additional requirements such as real-time analysis or extra hardware as required for active qubit reset.
	To achieve the highest restless repetition rate and SNR any given pulse sequence should begin immediately after the previous readout pulse.
	This should also reduce biases in the restless signal as the effect of $T_1$-decay will be minimized.
	In particular, RB measurements would not require post-selection and the effective repetition rate will be set by the gate fidelity as higher fidelity gates require longer Clifford sequences.
	The repetition rate is then limited by the duration of the pulse sequences, including the readout pulse, which can be shortened with optimal control \cite{Egger2014b, Bultink2016, Walter2017, Werninghaus2020} and may require reset after each measurement \cite{McClure2016, Bultink2016}.
	Future work would involve analyzing the effect of the higher excited transmon states \cite{Bianchetti2010} and demonstrating other forms of RB \cite{Magesan2012b, Morvan2020}.
	
	The higher repetition rates of restless measurements enable more frequent and significantly faster calibration and characterization.
	This would result in both extended up-times and/or more accurately calibrated systems.
	Furthermore, this may allow cloud-based quantum backends to deal with the calibration and characterization overhead required to implement Richardson Error mitigation \cite{Temme2017, Kandala2019}.
	
	\section{Acknowledgements}
	
	We thank F. Roy, S. Machnes, F. Wilhelm, M. Ganzhorn, M. Mergenthaler, P. Mueller, S. Paredes, M. Pechal, and G. Salis for insightful discussions as well as R. Heller and H. Steinauer for technical support. We also acknowledge useful discussions and the provision of qubit devices with the quantum team at IBM T. J. Watson Research Center, Yorktown Heights. This work was supported by the European Commission Marie Curie ETN QuSCo (Grant Nr. 765267), the IARPA LogiQ program under contract W911NF-16-1-0114-FE and the ARO under contract W911NF-14-1-0124.
	
	\appendix
	
	\section{Clustering single-shot data\label{sec:clustering_cost}}
	
\begin{figure}
		\centering
		\includegraphics[width=\columnwidth]{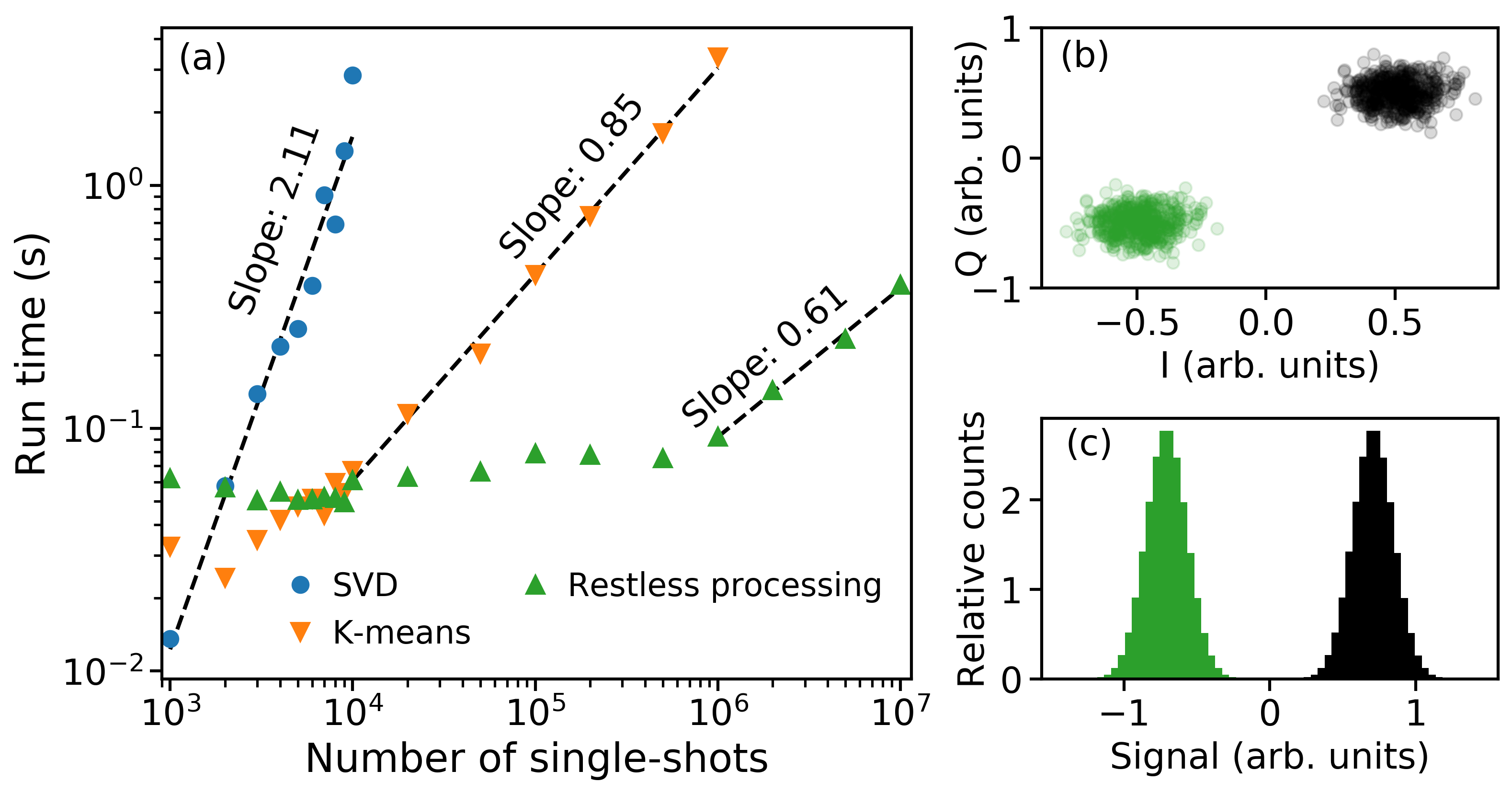}
		\caption{
			(a) Run time of SVD, $k$-means, and the restless analysis in Sec.~\ref{sec:restless_axis} as a function of the number of single-shots processed.
			The dashed lines show $\ln y=a\ln N_s +b$ fits used to obtain the scaling $a$ of the run-time.
			(b) The simulated IQ-data used, color-coded according to the $k$-means classifier.
			(c) The simulated IQ-data discriminated using the restless analysis.}
		\label{fig:run_time}
	\end{figure}

	Here, we compare the run time of SVD, $k$-means clustering \cite{Salman2011}, and the restless analysis in Sec.~\ref{sec:restless_axis} with a simplistic example.
	We numerically generate $N_s$ two-dimensional IQ-points forming two clusters centered around $(-0.5, -0.5)$ and $(0.5, 0.5)$ by sampling from a Gaussian distribution with a $0.2$ standard deviation, see Fig.~\ref{fig:run_time}(b).
	To ensure that the example includes the number of single-shots that we analyze in a typical experiment we vary $N_s$ from $10^3$ to $10^7$.
	The SVD (from numpy version 1.18.1) allows us to construct a signal axis, see Fig.~\ref{fig:axis_sketch}, while the $k$-means classifier (from sklearn version 0.21.3) with $k=2$ and the restless analysis allow us to build a state discriminator, see Fig.~\ref{fig:run_time}(c).
	The quadratic run time of SVD, see Fig.~\ref{fig:run_time}(a), makes it a poor choice when processing single-shot data.
	In our example, the $k$-means algorithm, based on Elkan's algorithm \cite{Elkan2003}, has an almost linear run time with the number of single-shots.
	Of the three methods the restless analysis performs best for $N_s>10^4$, see green triangles in Fig.~\ref{fig:run_time}(a).
	We expect a linear scaling $\mathcal{O}(N_s)$ as the number of samples increases.
	This simple example highlights that single-shots should not be analyzed with SVD when they cannot be averaged and that the restless analysis method is fast.

	\section{SPAM Errors}
	\label{sec:SPAM}
\begin{figure}[htbp!]
		\centering
		\includegraphics[width=\columnwidth]{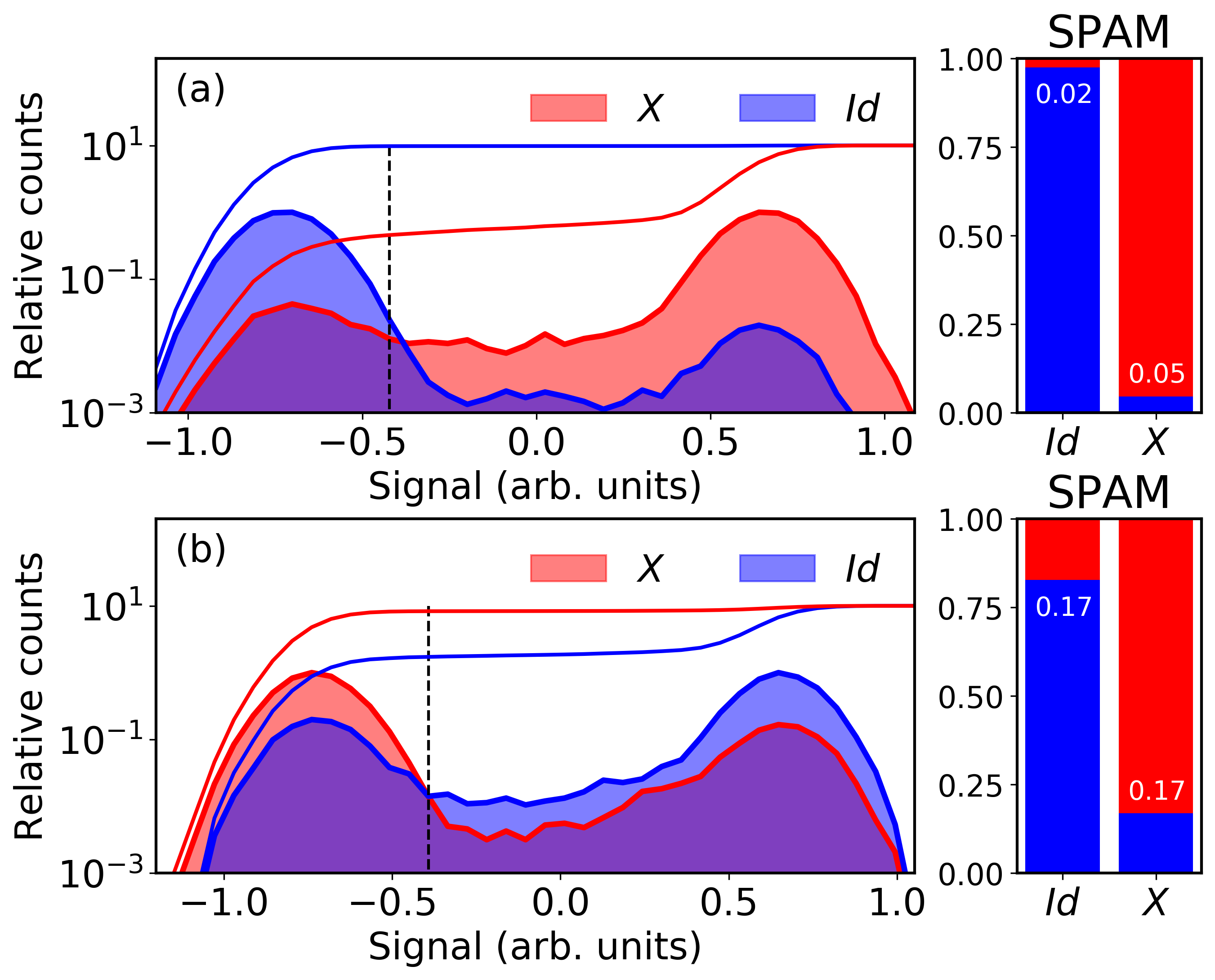}
		\caption{
			Histogram of restless measured IQ-points for $\mathcal{M}_A$ (a) and $\mathcal{M}_B$ (b) split into $Id$ (blue) and $X$ (red) operations.
			The blue and red solid lines show the cumulative distribution functions (scaled by a factor 10 for plotting) from which the discrimination threshold $x_T$, shown as a dashed black line, is obtained.
			The bar charts on the right show the SPAM errors $P_A(B|Id)=2\%$ and $P_A(A|X)=5\%$ in (a) and $P_B(A|Id)=17\%$ and $P_B(B|X)=17\%$ in (b).
		}
		\label{fig:meas_histo}
	\end{figure}

	The measurements described in section~\ref{sec:bias} allow us to quantify the SPAM errors for $\mathcal{M}_A$ and $\mathcal{M}_B$ by separating both data sets according to whether $G_k=Id$ or $G_k=X$, see Fig.~\ref{fig:meas_histo}(a) and (b), respectively.
	The single-shots $m_j$, projected onto the measurement axis form a one-dimensional distribution, see shaded curves in Fig.~\ref{fig:meas_histo}.
	From their cumulative distribution functions $F_{Id}$ and $F_X$, shown as solid lines in Fig.~\ref{fig:meas_histo}, we calculate an optimal threshold value for state discrimination given by the point where $F_{Id}$ and $F_X$ are maximally separated
	\begin{equation}
	x_T=\underset{x}{\arg\max} \left|F_X(x)-F_{Id}(x)\right|,
	\end{equation}
	shown as a black dashed line in Fig.~\ref{fig:meas_histo}.
	We use $x_T$ to discriminate the qubit states and extract the specific SPAM error $P_x(y|G)$ for $G\in\{Id, X\}$ and $x,y \in\{A, B\}$, see the SPAM plots in Fig.~\ref{fig:meas_histo}.
	
	\section{simulation of restless measurements\label{sec:restless_model}}
	
	Here we build a model of restless measurements to fit the fraction of points in $\mathcal{M}_A$ for the data shown in Fig.~\ref{fig:bias}. 
	We model the probability $p_{\ket{1},j}$ that measurement $j$ projects the qubit into the excited state, which depends on the previous measurements and operations, as
	\begin{align} \label{eqb:ma_model}
	p_{\ket{1},j} =& a\left[p_{\ket{1},j-1} + \eta_{k}\left(1-p_{\ket{1},j-1}\right) \right. \\ \notag
	&\left.-\eta_kp_{\ket{1},j-1}\right] e^{-1/R T_1}+b
	\end{align}
	Here, $\eta_{k}$ is the probability that $G_k$ flips the qubit state and $\exp{(-1/R T_1)}$ is the probability to decay due to $T_1$-relaxation in a time $1/R$.
	$a$ and $b$ are parameters to account for SPAM errors.
	The probability that the qubit was projected in the excited state during the previous measurement is $p_{\ket{1},j-1}$.
	The state transfer probabilities to and out of the excited state are $\eta_{k}(1-p_{\ket{1},j-1})$ and $\eta_kp_{\ket{1},j-1}$, respectively.
	
	We use the model in Eq.~(\ref{eqb:ma_model}) to fit $p_{k,A}$, experimentally estimated by $|\mathcal{M}_A|_k/N_s$, to
	\begin{align} \label{eqn:model}
	1-\frac{1}{N_s}\sum_{i=0}^{N_s-1}p_{\ket{1},k+iK}
	\end{align}
	which assumes $A=\ket{0}$, see Fig.~\ref{fig:bias}(a).
	We set $\eta_k=0$ and $\eta_k=99\%$ for $G_k=Id$ and $G_k=X$, respectively, and treat $T_1$, $a$, and $b$ as fit parameters.
	Since the qubit is in its ground state before the restless measurements begin $p_{\ket{1},0}=0$ which allows us to compute the sum in Eq.~(\ref{eqn:model}).
	This model accurately reproduces the data with fitted values $T_1=50.0(3.2)~\mu\rm{s}$ $a=0.983(26)$, and $b=0.084(12)$, see Fig.~\ref{fig:bias}(a).
	
	\bibliography{bibliography}
\end{document}